**Modelling the combined effects of land use and climatic changes: coupling bioclimatic modelling with markov-chain cellular automata in a case study in Cyprus**

Marianna Louca[1], Ioannis N. Vogiatzakis[1], and Aristides Moustakas[2,*]

1. School of Pure & Applied Sciences, Open University of Cyprus, Nicosia, Cyprus

2. School of Biological and Chemical Sciences, Queen Mary University of London, Mile End Road, E1 4NS, London, UK

* Corresponding author: Aristides (Aris) Moustakas

Email: arismoustakas@gmail.com

**Abstract**

Environmental change in terms of land use and climatic changes are posing a serious threat to species distributions and extinctions. Thus in order to predict and mitigate against the effects of environmental change both drivers should be accounted for. Two endemic plant species in the Mediterranean island of Cyprus, *Crocus cyprius* and *Ophrys kotschyi,* were used as a case study. We have coupled climate change scenarios, and land use change models with species distribution models. Future land use scenarios were modelled by initially calculating the rate of current land use changes between two time snapshots (2000 and 2006) on the island, and based on these transition probabilities markov-chain cellular automata were used to generate future land use changes for 2050. Climate change scenarios A1B, A2, B1 and B2A were derived from the IPCC reports. Species' climatic preferences were derived from their current distributions using classification trees while habitats preferences were derived from the Red Data Book of the Flora of Cyprus. A bioclimatic model for *Crocus cyprius* was built using mean temperature of wettest quarter, max temperature of warmest month and precipitation seasonality, while for *Ophrys kotchyi* the bioclimatic model was built using precipitation of wettest month, mean temperature of warmest quarter, isothermality, precipitation of coldest quarter, and annual precipitation. Sequentially, simulation scenarios were performed regarding future species distributions by accounting climate alone and both climate and land use changes. The distribution of the two species resulting from the bioclimatic models was then filtered by future land use changes, providing the species' projected potential distribution. The species' projected potential distribution varies depending on the type and scenario used, but many of both species' current sites/locations are projected to be outside their future potential distribution. Our results demonstrate the importance of including both land use and climatic changes in predictive species modeling.

**Introduction**

Climate and land use changes are two of the main causes for biodiversity loss worldwide, while their combined effects may be greater than either of these factors acting alone (de Chazal and Rounsevell, 2009). Climate change has already affected species distribution, abundance, phenology and interactions, while even greater impacts are expected in the future (Rosenzweig et al., 2007) with three major options for threatened species: (a) extinction, (b) evolution and subsequent adaptation and (c) shifting its geographic range to more favourable conditions (Moustakas and Evans, 2013). However, the rate at which climate change is happening today is often faster than the ability of some species to disperse or adapt, while other factors such as land use change and habitat fragmentation impede their ability to move to suitable areas (Thuiller et al., 2005).

Climate change models simulate the change in climate due to the accumulation of greenhouse gases, based on the current understanding of atmospheric physics and chemistry (Hannah, 2010). The Intergovernmental Panel on Climate Change (IPCC) has produced a range of emission scenarios for use in global climate models that predict future climate (IPCC, 2013). The latest IPCC report is based on alternative concentrations of greenhouse gases without being associated with any socio-economic scenario, but instead could result from different combinations of economic, technological, demographic, policy, and institutional futures (IPCC, 2013). This change facilitates better integration of socio-economic factors, such as land use changes into future projections.

Species Distribution Models (SDMs) use information on the locations of species and their corresponding environmental covariates, creating statistical functions to be projected in areas or time periods where environmental parameters are known but species distribution is unknown, providing inference for potentially suitable sites (Brotons et al., 2004).

In addition to climate change, the destruction, fragmentation and degradation of habitats due to changes in land use are among the strongest pressures on biodiversity (EEA, 2010). In analogy to climate modelling, land use change models use a variety of approaches to assess and project the future role of land use change on biodiversity, soil degradation, the ability of biological systems to support human needs and the vulnerability of places and people to climatic, economic, or sociopolitical perturbations (Zhang et al., 2014). The development of land use change scenarios allows their integration into SDMs alongside dynamic climatic variables which can significantly improve a model's explanatory and predictive ability at fine scales (Martin et al., 2013). Land use can be incorporated into the model as static variables that do not change over time (Iverson and Prasad, 2002), or as dynamic variables that change under different scenarios (Schweiger et al., 2012).

Despite the fact the combined effects of climate and land use change affect species distributions (Martin et al., 2013), only a small number of SDMs predict species distribution based on both of these factors (e.g. Esteve-Selma, et al., 2012; Schweiger, et al., 2012; Heubes, et al., 2013), while most SDMs that combine climatic variables with land use variables, only use dynamic variables for climate, while land use is considered stable (Martin et al., 2013). Dynamic model coupling (Verdin et al., 2014) of climate models and land use models can be employed to account for the interaction between both effects in projected future conditions (Evans et al., 2013a).

Cyprus is a biodiversity hotspot (Myers et al., 2000) which is expected to become warmer and dryer (Hadjinicolaou et al., 2011). At the same time, the increased pressure for urban and tourism development in Cyprus is leading to significant changes in land uses

(Eurostat, 2012). Thus, the island of Cyprus is an ideal study area because of the major climate and land use changes expected in the near future, the presence of a multitude of threatened endemic species and the absence of similar studies in the region to date. We sought to quantify the combined effects of climatic and land use changes on two plant endemic species.

**Materials and Methods**

### Study area
Cyprus is the third largest Mediterranean island, with an area of 9251 Km$^2$. The climate is Mediterranean, with hot and dry summers from June to September (little or no rainfall, average maximum temperatures up to 36°C), rainy but mild winters from November to March and two short transitional seasons, autumn and spring. For detailed information regarding the geomorphology and biogeography of the island see Supp. 1A.

### Target species
The target species are *Crocus cyprius* Boiss. & Kotschy and *Ophrys kotschyi* H. Fleischm. & Sofi, both endemic to Cyprus, and categorized as vulnerable under the IUCN classification (Tsintides et al., 2007). The criteria considered to target these two species were: (i) high risk of extinction (ii) endemism (iii) high number of data occurrences/location relative to other available species (iv) significant differences between their distributions, as *Crocus cyprius* only occurs in the Troodos Mountains, while *Ophrys kotschyi* occurs almost everywhere in Cyprus except the Troodos Mountains. For additional information regarding the target species see Table S1 in Supp. 1A.

### Data
Species distribution data were obtained from the Red Data Book of the Flora of Cyprus (Tsintides et al., 2007), in the form of true presence points. The data were collected during a systematic extensive survey between 2002 and 2006 (Tsintides at al., 2007).

There were 102 true presences for *C. cyprius* and 117 for *O. kotschyi*.. From these, a set of "theoretical" presences were derived for each species; these comprised centres of cells of the potential habitat, i.e. the entire area with a suitable altitude, soil and land use, within which at least one true presence had been recorded. Absence data were created artificially, from background data of the entire potential habitat of each species; "these comprised the centres of cells of the potential habitat where no true presence had been recorded. This was done in order to provide a sample of the set of conditions available to the species in the region and not to pretend that the species is absent in the selected sites (Phillips , et al., 2009). The data were then weighted to simulate prevalence 0.5, i.e. the total weight of presence is equal to the total weight of absences (Barbet-Massin et al., 2012).

Bioclimatic data were obtained from Worldclim database, Version 1.4 (release 3), which is available on [www.worldclim.org](http://www.worldclim.org) (Hijmans et al., 2005). The bioclimatic variables used are shown in Table S2 in Supp. 1B. Future bioclimatic data were also obtained from Worldclim for the year 2050, according to GCM HadCM3 (Hadley Centre Coupled Model, Version 3) and A1B, A2, B1 and B2A SRES emission scenarios. Each scenario is based on a different "storyline" and scenario family (A1, A2, B1 or B2), representing different demographic, social, economic, technological, and environmental developments (IPCC, 2013). The four storylines combine two sets of divergent tendencies: one set varying between strong economic values (A1 and A2 families) and strong environmental values (B1 and B2 families), the other set between increasing globalization (A1 and B1 families) and increasing regionalization (A2 and B2 families) (Nakicenovic & Swart, 2000).

. Land use data for 2000 and 2006 were obtained from CORINE database, available on http://www.eea.europa.eu/data-and-maps. The resolution for both the current and the future bioclimatic data was 0.71 $Km^2$ while for the land-use data 250 m x 250 m.

Habitat preferences were determined based on the information provided in The Red Data Book of the Flora of Cyprus, which was the result of a systematic study of all available information on the threatened plants of Cyprus, in combination with field work (Tsintides et al., 2007). The combination of suitable altitude, soil and land use, as described in Tsintides et al. (2007) was defined as the species current potential habitat when using current land use and as future potential habitat when using future land use in ArcGIS (http://www.esri.com).

**SDMs**

The SDMs were created using Classification Trees (CT), a machine learning method used to create predictive models (Figures 2 and 3). The method predicts the value of a dependent variable with a finite set of values, from the values of a set of independent variables (Ji et al., 2013). The main advantage of this method is that it does not require a specific type of data or that they follow a specific statistical distribution. The evaluation of the predictive accuracy of the model was measured using Cohen's Kappa (Congalton, 1991) and the "area under the curve" (AUC) of Receiver Operating Characteristic plot (ROC plot) (Fielding and Bell, 1997). The Classification Tree is considered to represent each species' "bioclimatic envelope" or "bioclimatic space", which is defined as the climatic component of the fundamental ecological niche, or the 'climatic niche' (Pearson and Dawson, 2003). We used SPSS version 20 for the statistical analysis (http://www-01.ibm.com/software/analytics/spss/).

**Land use prediction**

Future land use was predicted with an integration of Markov chain and Cellular Automata (Figure 4). Markov chain is a technique that has been widely used to predict changes in vegetation and which predicts future changes based on the rate of previous changes (Arsanjani et al., 2011). The main disadvantage of the Markov chain is its lack of spatial dimension: it gives accurate information on the transition probabilities of each land use type to another, but provides no information on the spatial distribution of changes (Eastman, 2003).

This problem is solved by combining Markov chain with Cellular Automata (Arsanjani et al., 2011). Cellular Automata are digital entities that have the ability to change their state based on the previous condition of themselves and their neighbours, based on a specific rule (Moustakas et al., 2006).

In order to model land use changes we used two different time snapshots 2000 and 2006 of the Cyprus CORINE land-use maps. For each land use type we calculated the transition probabilities to all the other land use types as well as the probability of no land use changes (retain current land use type). This was implemented in a spatially explicit manner i.e. transition probabilities were calculated for each location, where locations were represented by 1 km grid cells. Thus, the transition probability depends not only on the cell's previous state but also on the state and rate of change of the neighbouring cells.

### Climate and land use change integration

We generated future land use scenarios for 2050 using CA_MARKOV in Idrisi Selva software (http://www.clarklabs.org(Eastman, 2003) (Figure 4). We used ArcGIS version 8 (www.esri.com) to map the modelled outputs of the species' future potential distributions based on bioclimatic modelling (bioclimatic envelope and future climate data). We then combined the projected distribution for 2050 based on bioclimatic space with the current and future potential habitat, producing future potential distributions based on climate change only and on climate and land use changes respectively (Figure 5). This was done by simply taking the cells in which projected distribution based on bioclimatic space and potential habitat overlap. The methodology that was used is summarized in Figure 1.

**Results**
### SDMs

The potential distribution of *Crocus cyprius* can be predicted using only three bioclimatic variables as deduced from the classification tree: Mean Temperature of Wettest Quarter, Max Temperature of Warmest Month and, and Precipitation Seasonality; (Fig. 2). Model accuracy was "good" to "substantial" as indicated by Cohen's Kappa and AUC values of 0.719 and 0.859 respectively (Landis and Koch, 1977; Swets, 1988). The potential distribution of *Ophrys kotchyi*'s can be predicted using five bioclimatic variables as deduced from the classification tree: Precipitation of Wettest Month, Mean Temperature of Warmest Quarter, Isothermality, Precipitation of Coldest Quarter, and Annual Precipitation; (Fig. 3). Model accuracy was also good (Kappa = 0.619, AUC = 0.809). SDM performance measures only relate to current distributions.

### Land use prediction

The map of future land use produced from CORINE 2000 and 2006 maps using CA_MARKOV indicated that the biggest increase in 2050 compared to 2006 is predicted for artificial surfaces (Fig. 4). Non-irrigated arable land presents the greatest reduction, followed by sclerophyllous vegetation and natural grasslands (Fig. 4). The classes with the lowest predicted change are broad-leaved forest and mixed forest (Fig. 4).

### Climate and land use change integration

Potential distributions of *Crocus cyprius* in present conditions, considering climate change only and considering climate and land use change suggest that in scenario B2A its potential distribution increases, while in all other scenarios it decreases, with A1B being the most pessimistic (Fig. 5a; Table 1). The decreased potential distribution is limited at higher altitudes, while the suitable areas at lower altitudes disappear. Shifting is predicted in all scenarios, so that 34 to 68 out of the 102 species' current sites are excluded from its future potential distribution (Fig. 5a). The inclusion of land use change does not cause significant changes (Fig. 5a; Table 1).

In the case of *Ophrys kotschyi*, potential distribution is projected to decline in all scenarios, with B2A being the most optimistic and A2 the most pessimistic scenario (Fig. 5b, Table 2). The decrease is not limited to a specific area but is spread throughout the current potential distribution. Furthermore, shifting and fragmentation result in the exclusion of 111 to 116 out of the 117 species' current sites from its future potential distribution (Fig. 5b). The inclusion of land use change causes additional reduction, from 16.2% to 17%, resulting in the eradication of all of the species' current locations in scenario B1 and the persistence of one or two locations in all other scenarios (Fig. 5b, Table 2).

**Discussion**

According to our model all scenarios predict the disappearance of *Crocus cyprius*' main area of occurrence in the top of the Troodos Mountains, resulting in the extinction of many of its current sites. The extinction in this region is caused by the increase of the maximum temperature of the warmest month, which in all scenarios exceeds the upper limit of 25.68 ºC determined by the classification tree. The increased projected potential distribution does not necessarily imply that presence sites are safe (Supp. 1C and Table S3).

Model outputs for *Ophrys kotschyi* predict future locations of a species adapted to low precipitations and relatively high temperatures, but not to arid conditions – see Supp. 1C for more details. These conditions are found in the hilly areas and the central plain of Mesaoria, where this species occurs today. Therefore, the combination of decreased precipitation of wettest month, increased mean temperature of warmest quarter and decreased precipitation of coldest quarter in all scenarios is responsible for the reduction of the species' potential distribution (Table S4 in Supp. 1C). Scenario A2, which is the most pessimistic, predicted the greatest reduction in precipitation of wettest month, which is the most important predictor of the species distribution according to the classification tree. The loss, shifting and fragmentation of the *O. kotshyi*'s potential distribution are likely to be disastrous, since the vast majority of the species' present sites are not part of its future potential distribution in any scenario.

The inclusion of land use change plays a different role in each species. For *C. cyprius* it does not cause any significant changes compared to considering climate change only, while for *O. kotschyi* it is very important, causing an additional reduction in the species' potential distribution and in the number of present sites included in it. This difference is caused by the different requirements of each species regarding land use. For *C. cyprius,* suitable land uses consist only of coniferous forests, which in Cyprus mainly occur inside Natura protected areas and state land, thus they are not expected to undergo significant changes in the future. In contrast, suitable land uses for *O. kotschyi* consist of 10 different classes, the most important of which are non-irrigated arable land, sclerophyllous vegetation and natural grasslands. These three classes are expected to suffer the greatest reduction by 2050.

To our knowledge, no similar studies for the target species or species belonging to the same genera were found in the literature. However, the results of this study are consistent with other studies conducted in the Mediterranean region. (Esteve-Selma et al., 2012) predicted an increase or decrease of the potential distribution of the endemic *Tetraclinis articulata* in Southeast Spain, depending on the emissions scenario considered: The potential distribution increases in scenario B2 and severely reduced in scenario A2. This is in agreement with our results for *C. cyprius,* but not with those of *O. kotschyi*, whose potential distribution is predicted to decrease in all scenarios. Our results are also consistent with those of (Vennetier and Ripert, 2009), who predict the disappearance of most forest areas with high species richness in Southeast France by 2050, using bioclimatic modelling. The projected limitation of *C. cyprius*' potential distribution at higher altitudes agrees with the predictions of other studies (Bell et al., 2014; Ferrarini et al.). In general the coupling of climate change and land use change resulted in more restricted distributions for *O. kotschyi* is in agreement with the predictions of other studies (de Chazal and Rounsevell, 2009).

### Limitations, uncertainty and future directions

There are a number of limitations related with the methodology employed in the current study. These include spatial and temporal resolution mismatch between the bioclimatic and the species distribution data, the possibility of sampling bias on the species distribution data, the quality and accuracy of the Worldclim (Hijmans et al., 2005) and CORINE data (http://www.eea.europa.eu/data-and-maps). In addition SDMs are based on a number of unrealistic assumptions (Evans et al., 2013a). These include considering that the model quantifies the realized ecological niche (Pellissier et al., 2013) and that the species are in equilibrium with the environment, ignoring biotic interactions (Matias et al., 2014) and evolutionary and phenotypic changes as an adaptation to climate and land use change (Moustakas and Evans, 2013), assuming full dispersion for both species (Rodríguez-Rey et al., 2013) and training the model only in the realized environment (Maher et al., 2014). Also, it is considered that the potential distribution is the geographic area that meets one or more components of the fundamental ecological niche, when in fact the real potential distribution includes the entire fundamental ecological niche. As a result, the actual future potential distributions are likely to be under- or over- predicted (Jiménez-Valverde et al., 2008). Although widely employed as a measure for SDM performance, AUC has been also criticised (see Lobo et al. 2008). Other important sources of uncertainty were the creation of theoretical presences and absences (Barbet-Massin et al., 2012). In addition predictions are often scale specific while the interaction of species with their environment takes place at a variety of scales (Bellamy et al., 2013).

Finally the inclusion of both climatic and land use variables introduces a new source of uncertainty, through the different parameters and assumptions it brings (Conlisk et al., 2013). However simple models do not lead to generality in their predictions and thus increasing complexity may yield more realistic predictions (Evans et al., 2014; Evans et al., 2013b).

Here model coupling (Verdin et al., 2014) of bioclimatic modelling with the use of markov chain cellular automata (Arsanjani et al., 2011) was employed. Alternatives may include using individual based models (Gonzalès et al., 2013; Zhang et al., 2014) to model both land use & climatic changes with a presence only or presence absence model output and compare model outputs (Tonini et al., 2014) in an identical grid size. Although there is an interaction effect between land use and climate changes, disentangling these is conceptually complicated (see Lehsten et al. 2015) and is unaccounted for in this study.

The procedure developed herein can be used for any species where data is available and provides a valuable and transferable method for understanding potential shifts in species distributions. In addition the maps produced can guide actions on adaptation and mitigation measures to climate change for the species studied, providing a useful tool for policy and decision makers.

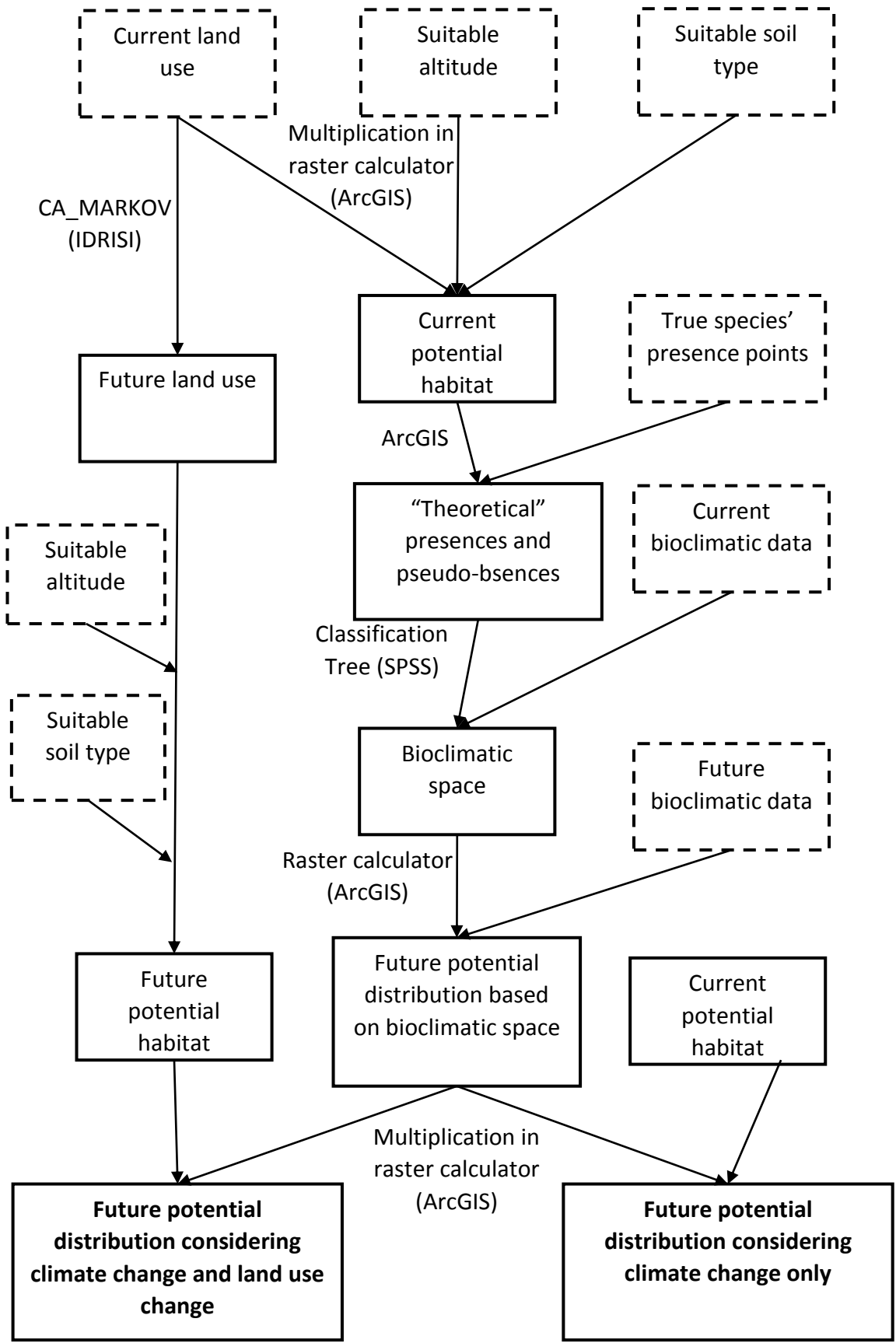

**Figure 1**: A graphical overview of the methodology. Data inputs are shown in dashed boxes, data outputs and intermediates in solid boxes and analysis methods with arrows.

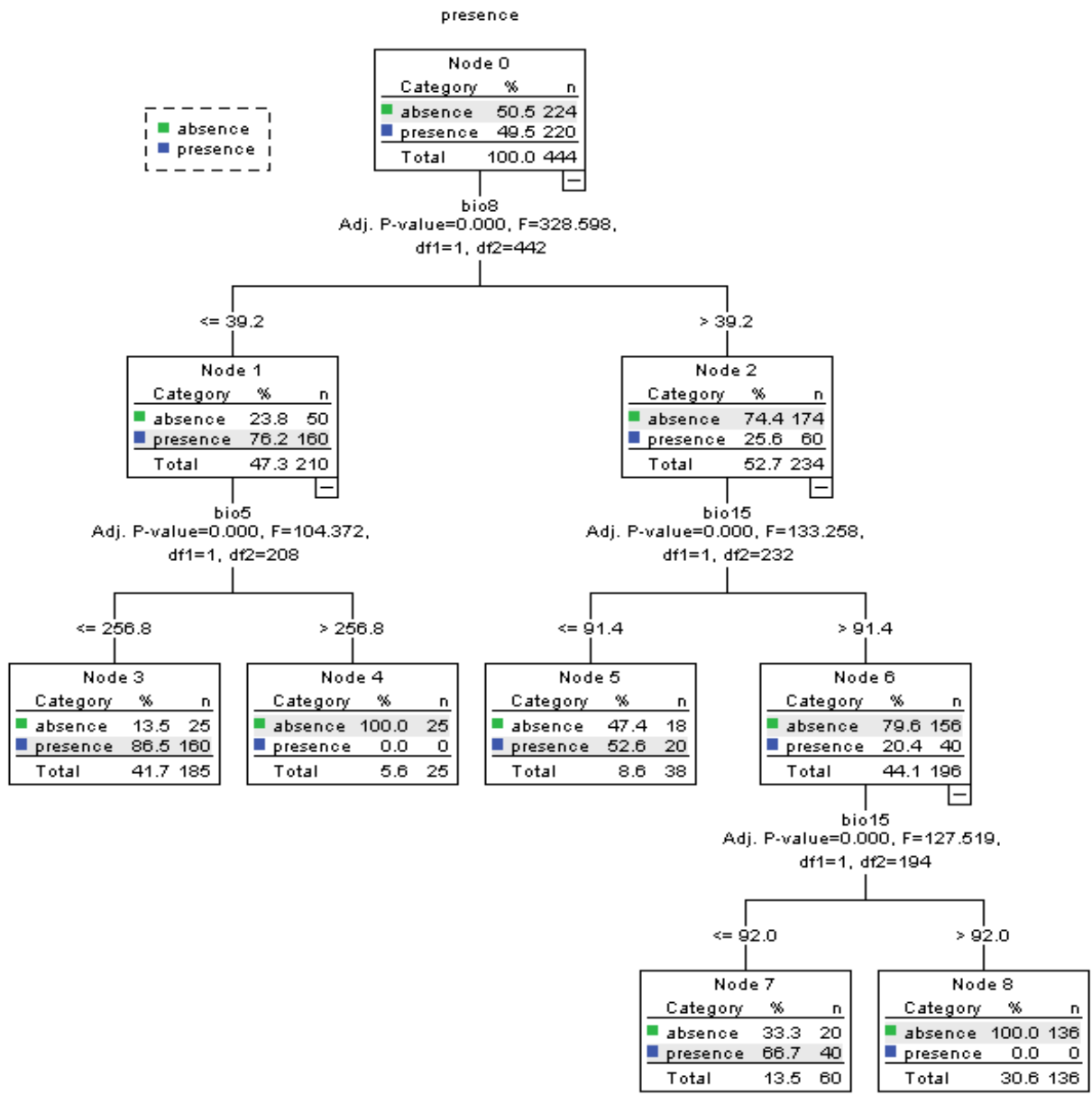

**Figure.2**. Classification tree for *Crocus cyprius*. As shown, three bioclimatic variables determine the distribution of *C. cyprius*: bio8 (Mean Temperature of Wettest Quarter), bio5 (Max Temperature of Warmest Month) and bio15 (Precipitation Seasonality). If bio8 is less than or equal to 3.92 °C, the species can only be present if bio5 is less or equal to 25.68 °C. If bio8 is more than 3.92 °C, the species can be present either when bio15 is less than or equal to 92%.

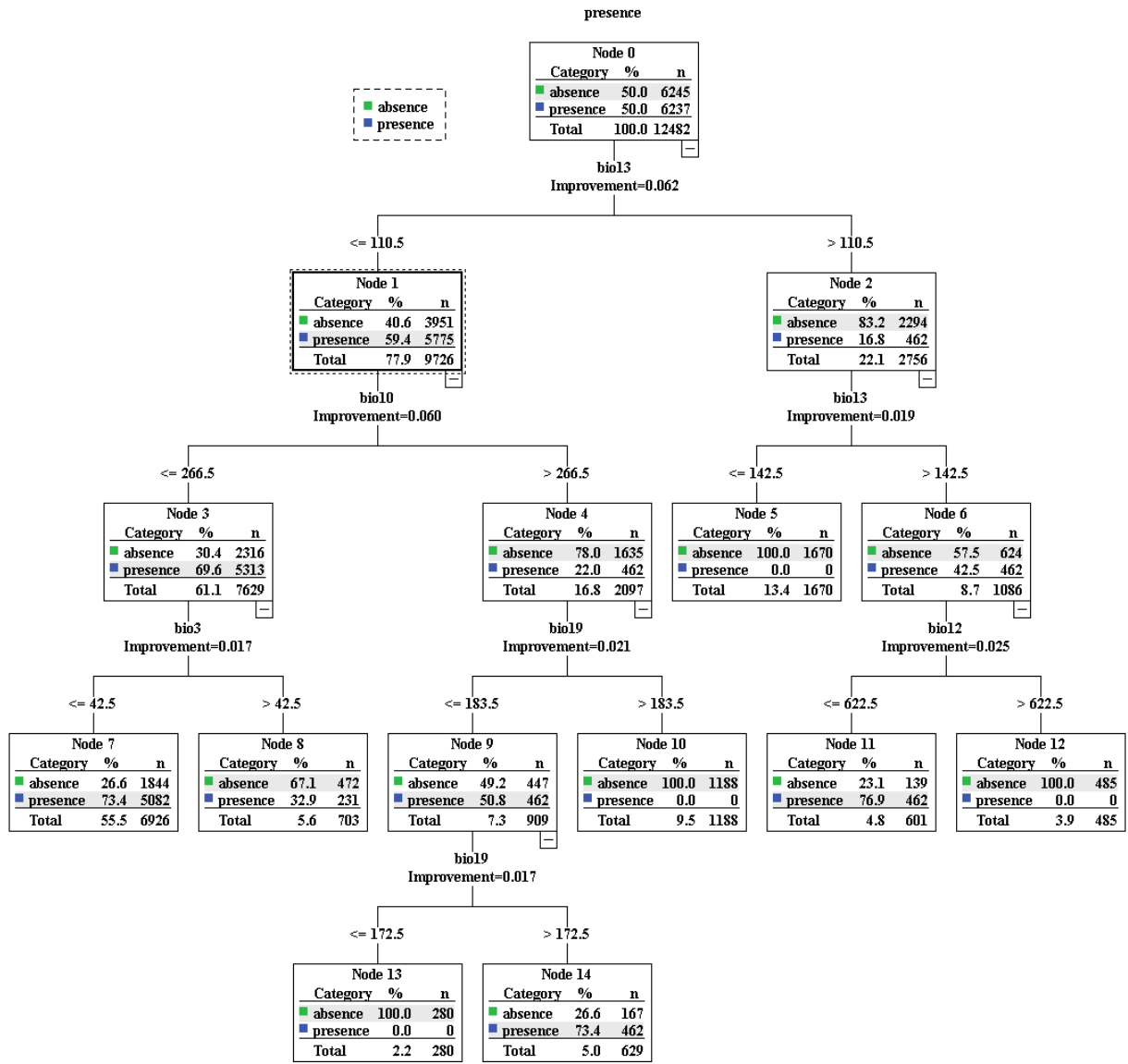

**Figure 3**. Classification tree for *Ophrys kotschyi*. A shown, five bioclimatic variables determine the distribution of *O. kotschyi*: bio13 (Precipitation of Wettest Month), bio10 (Mean Temperature of Warmest Quarter), bio3 (Isothermality), bio19 (Precipitation of Coldest Quarter) and bio12 (Annual Precipitation). If bio13 is less than or equal to 110.5 mm, the species can be present either when bio10 is less than or equal to 26.65 °C and bio3 is less than or equal to 42.5, or when bio10 is more than 26.65 °C and bio19 is between 172.5 mm and 183.5 mm. If bio13 is more than 110.5 mm, the species can only be present if bio13 is more than 142.5 mm and bio12 is less than or equal to 622.5 mm.

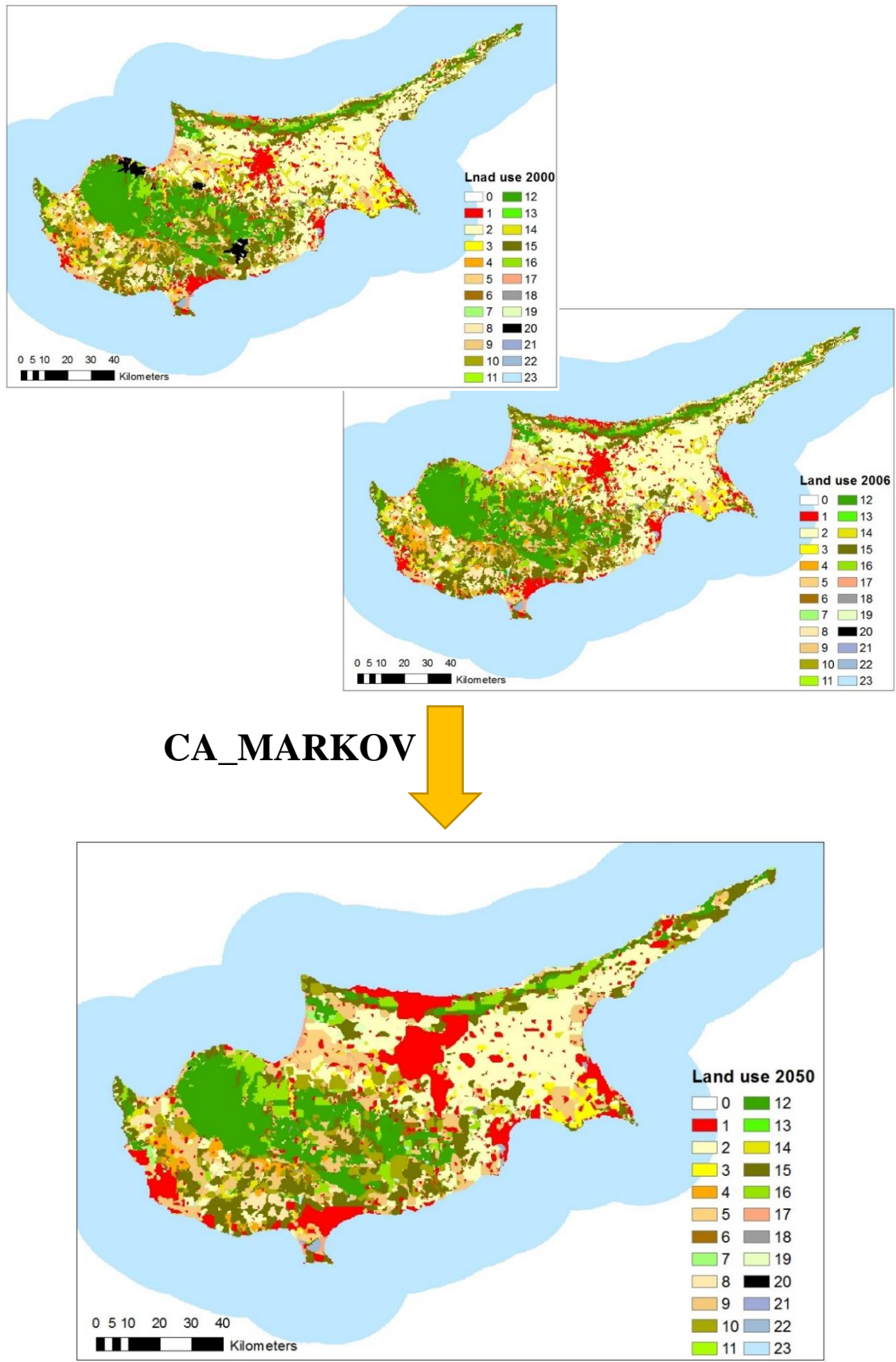

**Figure 4.** Prediction of land use in 2050, using CORINE land-cover data for 2000 and 2006 and CA-MARKOV. All land use classes and their codes are listed in Table S5 in Supp. 1.

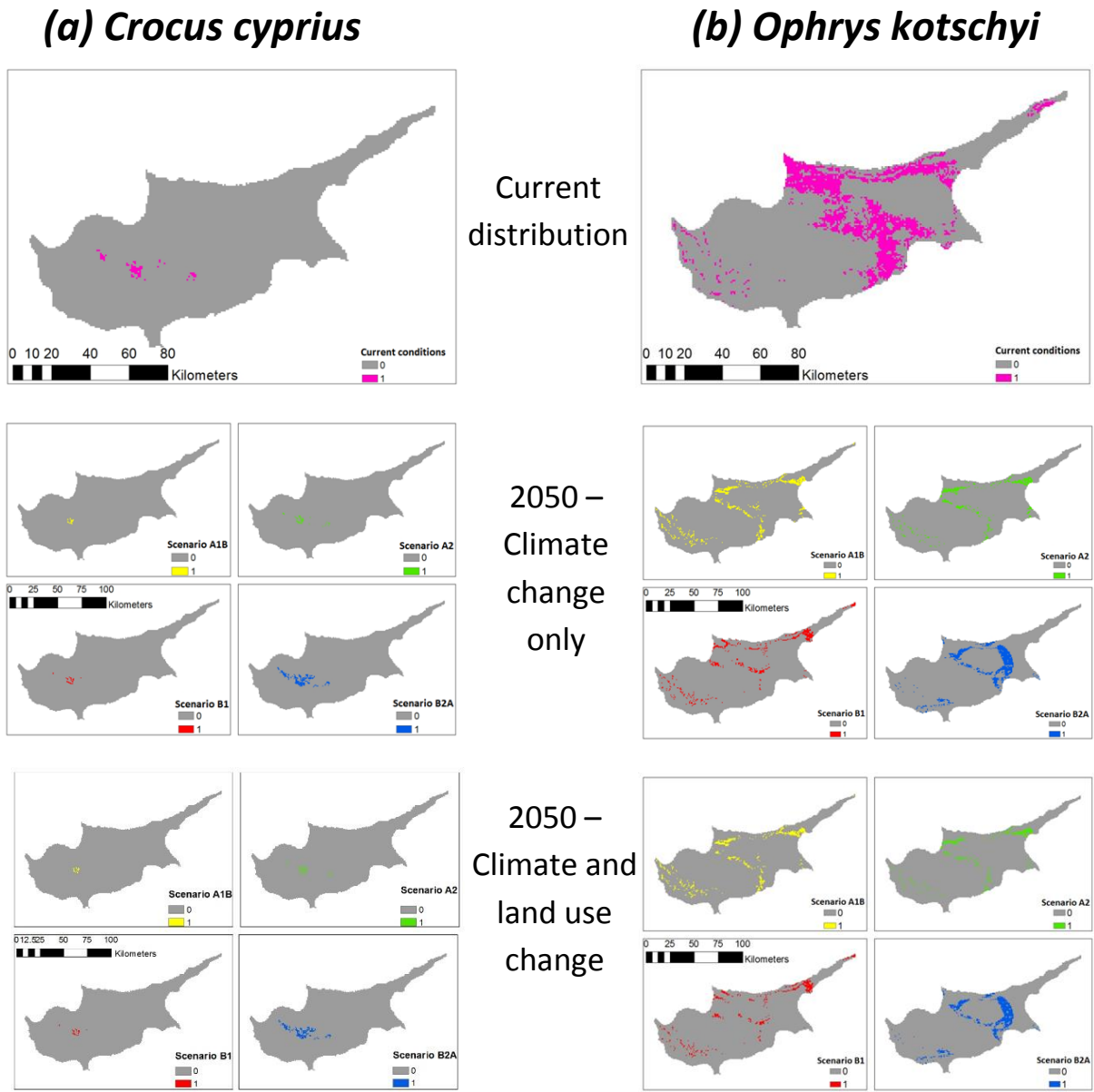

**Figure 5.** Potential distribution for *Crocus cyprius* (a) and *Ophrys kotschyi* (b) in current conditions (top), in 2050 considering only climate change (middle) and in 2050 considering climate and land use change (bottom). All land use classes and their codes are listed in Table S5 in Supp. 1.

**Table 1.** Number of cells and percentage cover of the potential distribution of *Crocus cyprius* compared to current potential distribution for each scenario used, considering climate change only and considering climate and land use change.

| Scenario | Number of cells – climate change only | Percentage cover – climate change only (%) | Number of cells – climate and land use change | Percentage cover – climate and land use change (%) |
|---|---|---|---|---|
| A1B | 22 | 31 | 22 | 31 |
| A2 | 53 | 74 | 52 | 72 |
| B1 | 31 | 43 | 31 | 43 |
| B2A | 212 | 294 | 204 | 283 |
| Current conditions | 72 | 100 | | |

**Table 2.** Number of cells and percentage cover of the potential distribution of *Ophrys kotschyi* compared to current potential distribution for each scenario used, considering climate change only and considering climate and land use change.

| Scenario | Number of cells – climate change only | Percentage cover – climate change only (%) | Number of cells – climate and land use change | Percentage cover – climate and land use change (%) |
|---|---|---|---|---|
| A1B | 690 | 31 | 573 | 26 |
| A2 | 580 | 26 | 479 | 22 |
| B1 | 569 | 26 | 477 | 22 |
| B2A | 974 | 44 | 906 | 41 |
| Current conditions | 2205 | 100 | | |

**Supplement 1**

**Supplement 1A**

The main geomorphologic zones of Cyprus are the Troodos Mountains in the southwest, the Pentadaktylos mountain range to the north, the central valley of Messaoria and the coastal zone. The main types of land cover include high forests on the Troodos and Pentadaktylos ranges, while the lower hills are dominated by shrubs alternating with built-up areas and cultivations. The plain of Messaoria and the coastal zone are mainly covered by cultivations and habitations, although some parts of natural or semi-natural vegetation still persist (Tsintides et al., 2007). Cyprus is predicted to be severely affected by climate change (Lelieveld et al., 2013). (Hadjinicolaou et al., 2011), predict the transition of the island to a warmer state for the period from 2026 to 2050, with increase in both the maximum and the minimum temperature by 1 °C to 2 °C and decline in rainfall by 8.2%.

**Table S1:** Target species information (IUCN, 2013; Tsintides et al., 2007). Both are included in Annex II of the Habitats Directive (92/43/EEC), and in Appendix I of the Bern Convention

| Species | *Crocus cyprius* | *Ophrys kotschyi* |
|---|---|---|
| Description | perennial herb | perennial orchid |
| Population | over 11,500 plants in the Troodos Mountains | at least 30 locations around the island usually forming small colonies of 10 to 100 plants |
| Soil type | igneous formations | limestone formations |
| Altitude | 1050 – 1950 m | 0 – 900 m |
| Land use type | pine forest openings, *Juniperus foetidissima* forests and edges of peat grasslands | phrygana and maquis, grassy slopes, field margins, sparse pine forests and moist places |
| Threats | trampling and construction, natural fires, climate change and military construction | land development, tourism infrastructure, overcollection and failure in sexual reproduction |
| Protection | all subpopulations within the Natura 2000 network | All subpopulations in state forests or in Natura 2000 areas |

-----------------------------------------------------------------------------------------------------------------

**Supplement 1B**

**Table S2.** Bioclimatic variables used in the model

| BIO1 | Annual Mean Temperature |
|---|---|

| BIO2 | Mean Diurnal Range (Mean of monthly (max temp - min temp)) |
|---|---|
| BIO3 | Isothermality (BIO2/BIO7) (* 100) |
| BIO4 | Temperature Seasonality (standard deviation *100) |
| BIO5 | Max Temperature of Warmest Month |
| BIO6 | Min Temperature of Coldest Month |
| BIO7 | Temperature Annual Range (BIO5-BIO6) |
| BIO8 | Mean Temperature of Wettest Quarter |
| BIO9 | Mean Temperature of Driest Quarter |
| BIO10 | Mean Temperature of Warmest Quarter |
| BIO11 | Mean Temperature of Coldest Quarter |
| BIO12 | Annual Precipitation |
| BIO13 | Precipitation of Wettest Month |
| BIO14 | Precipitation of Driest Month |
| BIO15 | Precipitation Seasonality (Coefficient of Variation) |
| BIO16 | Precipitation of Wettest Quarter |
| BIO17 | Precipitation of Driest Quarter |
| BIO18 | Precipitation of Warmest Quarter |
| BIO19 | Precipitation of Coldest Quarter |

**Supplement 1C**

*Crocus cyprius* can be found in areas that combine either low mean temperatures in the wet months (under 3.92 ºC), without extreme high summer temperatures (under 25.68 ºC) or high mean temperature of the wettest quarter (over 3,92º C) but relatively low precipitation seasonality (below 92%). Max Temperature of Warmest Month and Precipitation Seasonality are expected to increase in all scenarios, but A1B and B1 predicted the largest increase (Table S3). Additionally, Precipitation Seasonality is projected to increase in scenarios A1B and B1, while in A2 and B2A it remains at the same levels as today (Table S3). As a result, A1B has the most pessimistic projections in terms of future suitable area, while B2A the most optimistic (Table S3).

**Table S3.** Projected values of the bioclimatic variables that determine the distribution of *Crocus cyprius* in Cyprus (Worldclim, 2013).

|  | Max Temperature of Warmest Month (º C) | Mean Temperature of Wettest Quarter (º C) | Precipitation Seasonality (%) |
|---|---|---|---|

| **Present conditions** | 24.1-36.5 | 1.3-13.5 | 77-103 |
| --- | --- | --- | --- |
| **A1B** | 27.3-39.9 | 3.1-15.3 | 88-109 |
| **A2** | 26.9-39.4 | 2.9-15.1 | 77-105 |
| **B1** | 27.2-39.7 | 2.7-14.9 | 82-108 |
| **B2A** | 26.3-38.9 | 2.9-15.9 | 75-103 |

Increased projected potential distribution of *Crocus cyprius* does not necessarily imply that the species' present sites are safe. This becomes evident when comparing future potential distributions with the species' current true distribution. Although scenario A1B predicts the smallest potential distribution, most of the area where the species appears today is not affected and therefore does not affect the present sites. Contrastingly in scenario B1, although the potential distribution is predicted to be greater than that of A1B, most of the species' present sites are not included in it.

For *Ophrys kotschyi* the model determines that suitable areas combine: a) Low precipitation of the wettest month (under 110.5 mm), low mean temperature in the summer (under 26.65 $^0$C) and lower temperature diurnal range compared to annual range (isothermality under 42.5%), or b) Low precipitation of the wettest month (under 110.5 mm), high mean temperature in the summer (over 26.65 $^0$C) and low but not extreme low precipitation in the winter (between 172.5 and 183.5 mm), or c) Very high precipitation of the wettest month (over 142.5 mm) but low precipitation throughout the year (below 622.5 mm). See Table S4 below for more details.

**Table S4.** Projected values of the bioclimatic variables that determine the distribution of *Ophrys kotschyi* in Cyprus (Worldclim, 2013)

|  | Isothermality (%) | Mean Temperature of Warmest Quarter ($^0$ C) | Precipitation of Wettest Month (mm) | Precipitation of Coldest Quarter (mm) | Annual Precipitation (mm) |
| --- | --- | --- | --- | --- | --- |
| **Present conditions** | 29 - 46 | 18.1 - 27.2 | 69 - 234 | 152 - 622 | 326 - 1018 |
| **A1B** | 27 - 43 | 21.5 - 30.6 | 61 - 226 | 136 - 608 | 272 - 961 |
| **A2** | 28 - 44 | 20.8 - 29.9 | 47 - 213 | 132 - 581 | 260 - 947 |
| **B1** | 29 - 45 | 20.9 - 29.9 | 63 - 226 | 126 - 597 | 272 - 961 |
| **B2A** | 28 - 44 | 20.6 - 29.8 | 58 - 197 | 137 - 528 | 284 - 893 |

**Table S5.** Land use classes and their corresponding codes.

| Reclassification code | CLC CODE | LABEL 1 | LABEL 2 | LABEL 3 |
|---|---|---|---|---|
| 1 | | Artificial surfaces | - | - |
| 2 | 211 | Agricultural areas | Arable land | Non-irrigated arable land |
| 3 | 212 | Agricultural areas | Arable land | Permanently irrigated land |
| 4 | 221 | Agricultural areas | Permanent crops | Vineyards |
| 5 | 222 | Agricultural areas | Permanent crops | Fruit trees and berry plantations |
| 6 | 223 | Agricultural areas | Permanent crops | Olive groves |
| 7 | 231 | Agricultural areas | Pastures | Pastures |
| 8 | 241 | Agricultural areas | Heterogeneous agricultural areas | Annual crops associated with permanent crops |
| 9 | 242 | Agricultural areas | Heterogeneous agricultural areas | Complex cultivation patterns |
| 10 | 243 | Agricultural areas | Heterogeneous agricultural areas | Land principally occupied by agriculture, with significant areas of natural vegetation |
| 11 | 311 | Forest and semi natural areas | Forests | Broad-leaved forest |
| 12 | 312 | Forest and semi natural areas | Forests | Coniferous forest |
| 13 | 313 | Forest and semi natural areas | Forests | Mixed forest |
| 14 | 321 | Forest and semi natural areas | Scrub and/or herbaceous vegetation associations | Natural grasslands |
| 15 | 323 | Forest and semi natural areas | Scrub and/or herbaceous vegetation associations | Sclerophyllous vegetation |
| 16 | 324 | Forest and semi natural areas | Scrub and/or herbaceous vegetation associations | Transitional woodland-shrub |
| 17 | 331 | Forest and semi natural areas | Open spaces with little or no vegetation | Beaches, dunes, sands |
| 18 | 332 | Forest and semi natural areas | Open spaces with little or no vegetation | Bare rocks |
| **19** | 333 | | Open spaces with little or no vegetation | Sparsely vegetated areas |
| **20** | 334 | | Open spaces with little or no vegetation | Burnt areas |
| **21** | 411 | Wetlands | Inland wetlands | Inland marshes |
| **22** | 421 | | Maritime wetlands | Salt marshes |
| **23** | | Water bodies | - | - |